# Maximizing Utilization and Performance of Guaranteed-Bandwidth Long Fat Networks and Virtual Circuits

D. Michael Freemon
National Center for Supercomputing Applications
University of Illinois
Urbana, IL, USA
mfreemon@illinois.edu

*Abstract*—Like many big science projects, the Large Synoptic Survey Telescope (LSST) has multiple geographic locations among which large amounts of data must be transferred. One particular type of data, crosstalk-corrected images, must be moved from South America to North America under stringent deadline requirements. LSST is provisioning an international network with bandwidth guarantees to handle this traffic. In prior work, we re-examined TCP congestion control for this use case and found that TCP throughput can approach wire speeds. This work shows that the Hierarchical Token Bucket (HTB) provides an excellent mechanism by which bandwidth can be managed for a wide range of traffic types. Using HTB without TCP congestion control over guaranteed-bandwidth virtual circuits is a compelling solution to the historical problem of poor TCP performance over long fat networks.

## I. INTRODUCTION

The Large Synoptic Survey Telescope (LSST) is a large, ground-based astronomical facility with an observatory in Chile and a data processing and archive facility located in the United States. There are stringent deadlines by which certain image files must be transferred from South America to North America. The network design for the project provisions a guaranteed 10 Gbps of bandwidth between La Serena, Chile, and Champaign, IL, US. Prior work[1] demonstrated that the data can be transferred over the high latency network at wire speed by disabling TCP congestion control.

The contribution of this work addresses the application controller concept left unresolved from the prior work[1], implements Linux kernel modifications to constrain the retransmission tail discussed in that paper, incorporates the latest LSST system design decisions taken since that paper was published, takes into account more types of network traffic that have different requirements and characteristics, suggests a mechanism to implement project policy regarding the use of the network, and increases the throughput due to improvements to the Linux kernel modifications used to disabled TCP congestion control.

We find that the Hierarchical Token Bucket (HTB)[4] is an effective solution for managing the bandwidth. Since the filtering capabilities of the Linux traffic control software[5] can distinguish among the different types of LSST traffic, HTB is sufficient for fulfilling the application controller role envisioned in Freemon[1].

The combination of disabling TCP congestion control, the use of HTB to manage the bandwidth, and leveraging virtual circuits is a compelling solution to the general problem of throughput on long fat networks.

The remainder of this paper is organized as follows. Related work is summarized in Section II. Section III outlines the requirements, design, and implementation of the proposed solution. In Section IV, the benchmark testing is described and the results are presented. Section V is a discussion of related topics. In Section VI, future work is suggested. Section VII concludes.

## II. RELATED WORK

The prior work, Freemon[1], demonstrated that disabling TCP congestion control on guaranteed-bandwidth networks does solve the throughput issues that exist for traditional shared-bandwidth networks. It also addressed a number of concerns that arise when disabling congestion control. Managing the priority and amount of data placed on the network becomes the responsibility of the application, not the network. Various implementation approaches were discussed in general terms, but was silent regarding specific guidance.

Exponential backoff of the retransmission timer was introduced by Jacobson[11] as part of the feature set added to the TCP protocol standards to avoid congestive collapse of the Internet. His work assumed shared-bandwidth networks, which does not apply in our case.

The idea of reducing the exponential backoff of the retransmission timer is not new. Mondal[12] shows that exponential backoff can be removed and, further, that its use is inappropriate in modern networking environments.

Vasudevan[14] shows that eliminating the minimum bound for the retransmission timer solves the problem of TCP incast collapse in data center networks, where round-trip latency is typically well under 1ms. They also show that it is safe to use reduced retransmission timeout values in wide area networks.

Much of the recent work reviewing the appropriateness of the traditional TCP congestion control features has arisen as a result of wireless networks, which, as in our case, also experience packet loss that is not indicative of network congestion. Caceres[13] "forces" the fast retransmission mechanism when packet loss is not caused by congestion, thereby avoiding the long delays caused by retransmission timeouts.

Dong[17] proposes a new approach to congestion control that does not use packet loss to directly control the sender's rate of packet transmission, as standard TCP does. They reinforce the view that packet loss events do not necessarily correspond with network congestion, and introduce a new learning algorithm that aggregates the signals from various network events to adjust the sending rate to meet its performance objectives.

Zurawski[6] uses Linux traffic control and HTB over OSCARS[7] virtual circuits to address out-of-order and burst maladies associated with QoS implementations in OSCARS. Although we previously addressed[1] burst issues and will not experience the multiple ingress queue problem, Zurawski does provide another concrete example of the use of HTB over virtual circuits. Zurawski states that this approach "make[s] it possible to use a high percentage (90-92% in practice) of bandwidth reservations". That number is for large bulk transfers with normal congestion control (slow start, etc.), so it is not the same as the LSST requirement for crosstalk-corrected images, but it is indicative of similar approaches and similar performance results.

| Data Type | Description |
|---|---|
| Crosstalk-corrected Images | This data must get transferred as quickly as possible. A burst of data every ~17 seconds. |
| Raw Images | The same amount of data and cadence as the crosstalk images. This is not time sensitive. |
| EFD | Database replication. This data needs to move at a constant rate, unaffected by the other traffic on the network, but at a lower bandwidth. |
| Catchup Images | This is the backlog of images that have piled up at La Serena after an outage of some kind. This is not time sensitive. |
| Interactive | Unscheduled traffic. Very small amounts of high priority traffic (because user response time depends on it), such as ssh sessions. |
| Ad hoc | Unscheduled traffic, such as user-initiated file transfers. |

Table 1. Types of network traffic in the LSST Data Management System.

Liu[16] looked at the suitability of using rate-guaranteed virtual circuits instead of normal IP-routed networking for file transfers. Liu envisions short-lived virtual circuits, which is not necessarily the LSST scenario, and Liu does not use modified TCP congestion control. As a result, their maximum observed throughput is still significantly less than the bandwidth capacity provided by the network, and they still require parallel TCP streams to reach their maximum. Liu uses virtual circuits only to separate alpha flows from other network flows.

McGinley[18] also examines the use of rate-guaranteed virtual circuits for file transfers. The author defines three types of virtual circuits, of which only his "Type 2" is applicable to us (because we do shaping at the sender). For Type 2, McGinley suggests the use of the CTCP[19] congestion control algorithm. CTCP is not sufficient[20] to meet our requirements[1]. Neither is TCP Illinois[21], Scalable TCP [22], HSTCP[23], nor any of the other existing variations[24], because they all react to packet loss by reducing transmission rates.

McGinley argues that a new transport protocol is needed for use with Type 1 and Type 3 virtual circuits, on which we do not comment further, as those types do not apply in our scenario. As an aside, we believe that the use of source-based traffic control should not be considered part of the virtual circuit definition, as the virtual circuit implementation should be independent of how the sender decides to shape and/or schedule its packets for transmission.

| Data Type | Priority | Guaranteed Bandwidth | Maximum Bandwidth | Unused Bandwidth |
|---|---|---|---|---|
| Crosstalk-corrected Images | 0 | 95% | 100% | Can be used by anyone else |
| EFD | 1 | 5% | 5% | Can be used by anyone else |
| Interactive | 2 | ~0% | 10% | Can be used by anyone else |
| Raw Images | 3 | ~0% | 100% | Can be used by anyone else |
| Catchup Images | 4 | ~0% | 100% | Can be used by anyone else |
| Ad hoc | 5 | ~0% | 100% | Can be used by anyone else |

Table 2. Bandwidth policy for the LSST Data Management System. Priority zero is the highest. The percentages are based on the total available bandwidth. The references to "~0%" are actually given 1 Mbps of guaranteed bandwidth to prevent TCP session failures.

III. REQUIREMENTS, DESIGN, AND IMPLEMENTATION

A. *Available Bandwidth*

The LSST system design has 21 "replicator" nodes in South America and 21 "distributor" nodes in North America to be used for the network traffic. These nodes are standard Linux servers and operate in pairs, with one replicator node sending data to its corresponding distributor node. The network between these sites has a guaranteed 10 Gbps of bandwidth. This means that each replicator/distributor pair has 476 Mbps of bandwidth. The amount of crosstalk-corrected image data that is sent across a single replicator/distributor pair is 128 MB. The project has placed a 5-second deadline by which that file transfer needs to be completed.

## B. Types of Network Traffic

There are several types of network traffic that need to be handled correctly by the bandwidth management solution. They are listed in Table 1. The "EFD" acronym is a reference to the LSST Engineering and Facility Database, which must be replicated between the two sites.

## C. Traffic Policy

The traffic policy is defined by the project according to the project's requirements. These are the rules that the bandwidth management solution needs to enforce. They are given in Table 2.

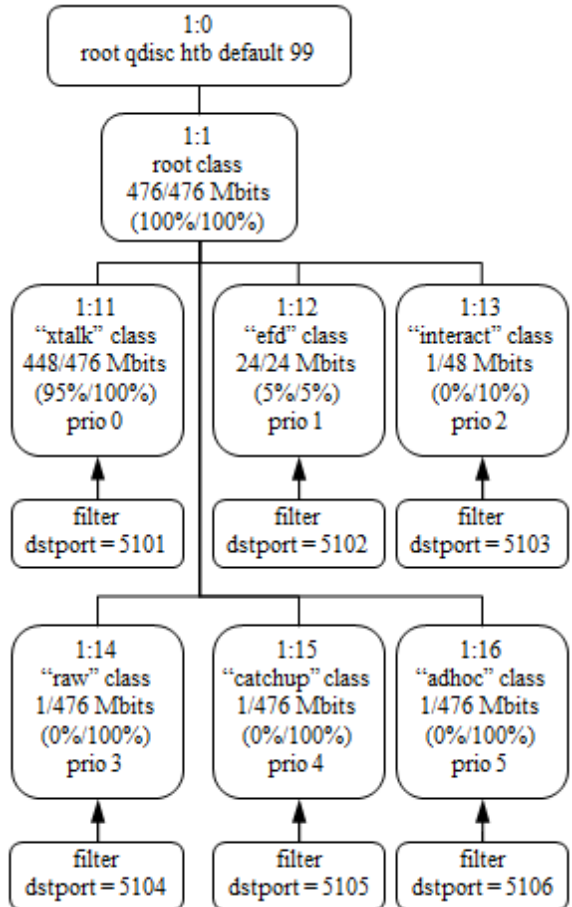

Figure 1. The Hierarchical Token Bucket (HTB) classes and filters used to implement the bandwidth policy given in Table 2.

In Table 2, for each type of data, the guaranteed bandwidth is available to that data type if needed, but under no circumstances will it exceed the maximum bandwidth. Priority 0 is the highest, and higher priority types get any available bandwidth in preference to lower priority types, within the constraints given by the guarantee and maximum. These percentages are relative to the total amount available to the node. For this scenario, the total amount available to the node is 476 Mbps. The data types with an indicated "~0%" of guaranteed bandwidth are actually given 1 Mbps of bandwidth to prevent TCP session failures.

## D. Bandwidth Management

Prior work[1] suggested a generic application controller daemon be responsible for actively managing the bandwidth allocations among the TCP senders in real-time, based on application events, such as notification of a new file that needs to be transferred. While that is indeed the more general solution for the widest possible range of requirements, any particular system could look to simplify that design based upon its specific requirements. The LSST policies and requirements are just such a case. Since we can distinguish the different traffic types based upon IP addresses and ports, we can use the Hierarchical Token Bucket (HTB)[4] queue discipline to implement the bandwidth management. This approach avoids the work, complexity, and failure modes of writing and using custom code.

Figure 1 gives the HTB class definitions and filters in effect during the benchmark testing. The bandwidth specifications are line rate allocations. The first number (before the forward slash) is the guaranteed bandwidth for that class, the second number is the maximum bandwidth. Any unused bandwidth is shared among all the other classes in priority order. This is the implementation of the policy given in Table 2.

## E. Disabling TCP Congestion Control

The prior work[1] disabled TCP congestion control by implementing a new Linux kernel congestion control module. Although that was effective, there were still conditions under which the kernel would reduce the congestion window despite the values provided by the kernel module. The current work takes a different approach by simply modifying the kernel code itself. This is straightforward code, and testing showed this to be a superior implementation, in that the congestion window never wavered from the desired setting.

Not all TCP sessions are treated in the same way. Specifically, only those connections going to a specific IP address range and port range have their congestion window disabled. Congestion control for all other traffic is still used in the normal way.

Disabling congestion control was implemented by setting cwnd to 99999, which is higher than the maximum value that would normally be encountered for these flows. The initial congestion window (INITCWND) was also set to 99999.

## F. Accelerating Tail Retransmissions

The file transfers for the crosstalk-corrected images must complete as soon as possible. When packet loss occurs, the data affected must be retransmitted. TCP sets a retransmission timer (RTO) to fire in the event that previously transmitted data has not been acknowledged by the receiver within a certain amount of time. The calculation for setting that timer is specified in RFC 2988[8], which includes exponential backoff of the retransmission timer. Exponential backoff is the doubling of the timer delay each time a retransmission timeout occurs. The effect of this can be seen in Figure 2.

The considerations that led to the development of the exponential backoff rule[11] do not apply here as we have a bandwidth guarantee. Thin streams[9][15], and in particular its linear backoff component, are a step in the right direction, but we can do better than linear, as we are willing to accept the tradeoff of a small number of spurious retransmissions in order to get the final bytes to the destination as quickly as possible.

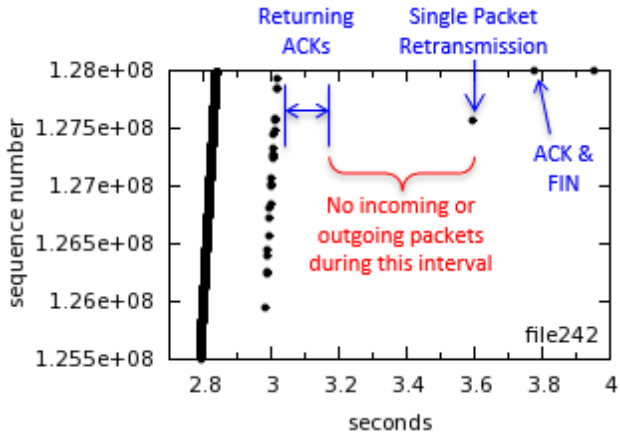

Figure 2. Exponential backoff of the TCP retransmission timer. This is a sequence diagram of the final portion of a transfer with 180ms latency and 1% packet loss. The solid line on the left is the first transmission of the data, the "dotted line" shows retransmitted packets due to loss. The single packet retransmission at second 3.6 is the second retransmission of that particular packet. Note the large interval of time between 3.2 and 3.6 where no activity at all was occurring on the network. During this interval, the kernel has no way to trigger fast retransmit, requiring it to wait for the RTO timer to fire. Because it has already retransmitted once, exponential backoff has doubled the normal wait time.

We introduce the following modifications to the RTO timer calculation:
- If the number of packets in flight is less than 100, then set the RTO timer to 50 milliseconds
- If the number of packets in flight is less than 6800, then set the RTO timer to 185 milliseconds
- Otherwise, leave the RTO timer unmodified from the value set by the normal algorithm

This has the effect of (a) using a linear timeout during the last RTT of the transfer since packets in flight will be approximately 6960 when the pipe is full and sending at the full rate, and (b) retransmitting lost packets in the tail every 50ms until they are acknowledged. A few of these packets will turn out to be spurious retransmissions, but that does not cause any problems.

## IV. BENCHMARK TESTING

### A. Equipment Setup

Two Linux machines serve as the TCP endpoints, with a third Linux machine between them acting as a router. The router machine injects latency and packet loss to simulate the long fat network.

The sending machine is an Intel i7-4770 @ 3.4 GHz with 16GB DDR3-1600 RAM running Fedora 20, kernel 3.15.10-200.fc20.x86_64 with custom modifications to disable TCP congestion control and accelerate tail retransmissions (discussed in this paper). The receiving machine is an Intel 6600 @ 2.4 GHz with 2GB DDR2-800 RAM running Fedora 20, kernel 3.15.10-200.fc20.x86_64. The intermediate router machine is an Intel Pentium 4 Extreme Edition @ 3.73GHz with 4GB DDR-667 RAM running Fedora 19, kernel 3.12.6-200.fc19.x86_64.

The network interface cards are Intel Gigabit CT Desktop 82574L PCIe x1 adapters with an MTU of 1500 bytes. TSO, GSO, and GRO are all disabled on all machines. The network switch is an entry-level consumer-grade Trendnet TEG-S50g with a 104K buffer.

Netem[2] is used to simulate the wide-area network. For latency and packet loss, we use the same assumptions and implementation as in the prior work[1]. Normal TCP/IP tuning (buffers, window sizing, etc.) was applied on all servers. Additionally, the sending machine had its NIC's rx ring set to 512, and the initial receive window on the receiving system was set to 99999 using the ip route command. The nuttcp program[3], version 7.1.6, was used to generate all network traffic.

### B. Timing of Traffic Events

During the benchmark testing, various flows are started and allowed to complete in order to measure the effectiveness of HTB at adjusting bandwidth allocations in real time as demand changes. Table 3 provides the approximate times at which the indicated events occur. All flows send as much data as possible while they are active.

| Time (sec) | Event | Time (sec) | Event |
|---|---|---|---|
| 0.4 | EFD starts | 17.2 | raw ends |
| 3.4 | adhoc starts | 19.2 | catchup ends |
| 6.4 | catchup starts | 21.4 | xtalk starts interactive starts |
| 9.4 | interactive starts | 23.2 | raw starts |
| 12.4 | xtalk starts | 23.8 | xtalk ends |
| 14.2 | raw starts | 26.4 | raw ends |
| 14.8 | xtalk ends | 27.5 | interactive ends |
| 15.5 | interactive ends | | |

Table 3. Timeline of events during the benchmark runs. The events mark the start or end of the transmission of data of the given type. These data types are defined in Table 1.

### C. Test Results

Figure 3 shows the test results. There are three plots, corresponding to 0% packet loss, 0.01% loss, and 1% loss, all using 180ms latency. Test runs were also conducted at 0.001% and 0.1% packet loss levels, but those results were consistent with the others, so in the interest of space are not shown in Figure 3. Utilization of the available network bandwidth is nearly 100% during the entire interval, and packet loss has no significant effect on throughput. Scanning across Figure 3 and comparing with the events described in Table 3, we see that HTB is enforcing the bandwidth policies exactly as desired.

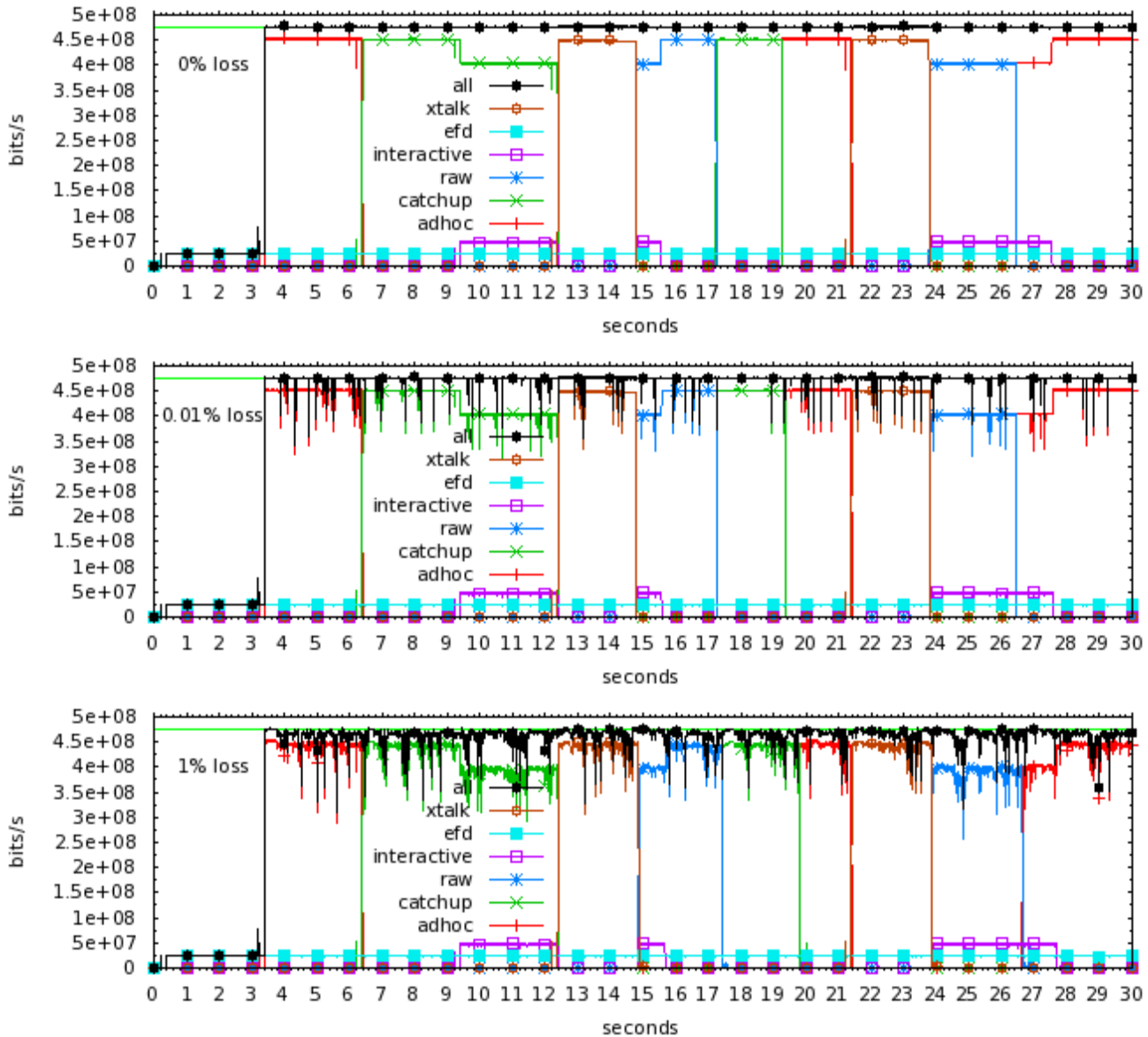

Figure 3. Linear graphs of the test results. All tests use 180ms latency (RTT). The top plot is 0% packet loss, the middle plot is 0.01% packet loss, and the bottom plot is 1% packet loss. Utilization is not affected by packet loss, and throughput is only trivially affected, as we show in the crosstalk-corrected image transfer section of this paper. We conclude that the apparent drops in utilization, which is most noticeable on the bottom plot, is an exaggerated visual artifact caused by the plot itself, since the throughput times are only trivially affected by the packet loss (see Table 6 and Table 7).

Figure 4 depicts a selected portion of the data from the 0.01% packet loss test. The vertical axis is using a log scale, which highlights the slower data flows. We again see what we expect. The lower priority flows are limited to just 1 Mbps during periods in which higher priority data classes are active.

### D. Accelerated Tail Retransmissions

Figure 5 shows the results of the new retransmission algorithm. These are plots from some of the worst performing transfers at 1% packet loss, intentionally selected to show the retransmission patterns. We see confirmation that the intended behavior is working, thus constraining the length of time that a transfer will take to complete.

Table 4 and Table 5 is the statistical comparison of the default Linux retransmission timer algorithm (exponential backoff) with the modified timer algorithm. It shows that, at 1% packet loss, the maximum time that the crosstalk transfers will take is 0.83 seconds faster with the modified algorithm.

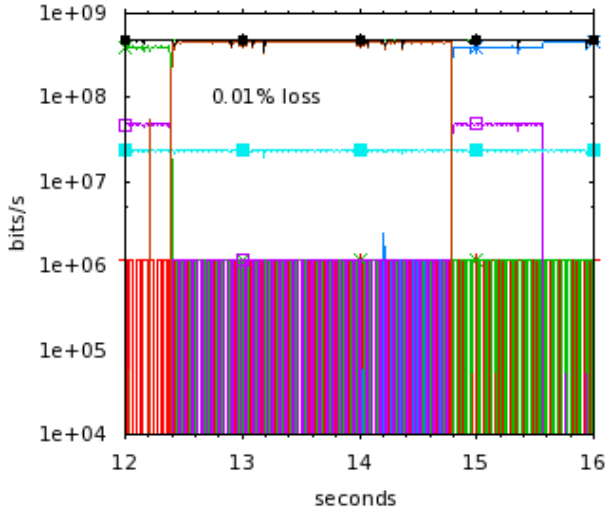

Figure 4. A log scale plot of a selected range of the data from the 0.01% packet loss test. The brown line (hollow box with tics) is a crosstalk-corrected transfer, which dominates all other traffic except for the EFD data (cyan/solid box). This plot reflects a correct implementation of the desired policy.

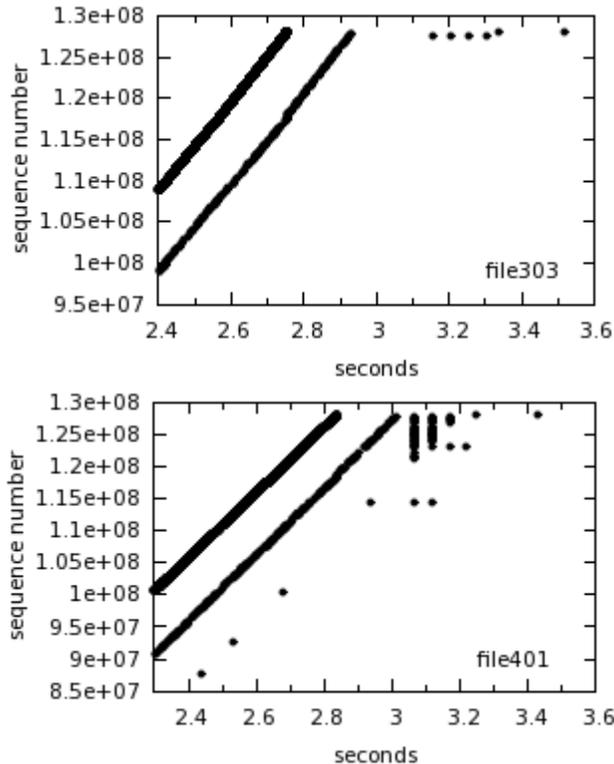

Figure 5. Accelerated Tail Retransmissions. Both of these are 180ms latency (RTT) with 1% packet loss. In the top case, one of the last packets in the stream was lost twice. This is the same case that we saw in Figure 2, except that now we retransmit after a single RTT, and continue to retransmit every 50ms until the receiver acknowledges the packet. In the bottom plot, we see a particularly pathological case, where a number of packets (or their returning ACK packets) at the end of the file transfer are lost. The lost data is retransmitted every 50ms, thus ensuring a timely completion to the file transfer despite the harsh conditions on the network.

We did compare Table 4 with the prior work[1], and found that the data matched up well, with the one exception of the mean value at 1% loss. The prior work had that value at 3.04 seconds and this paper shows it to be 3.25 seconds (note that the prior paper inadvertently had the mean and the median values reversed). A number of factors make this apparent discrepancy difficult to resolve, including different hardware, faster CPUs, more and faster RAM, updated software on all systems including kernels (3.15.10 vs 3.7.9), and different bandwidths in use (1 Gbps vs 476 Mbps, both on 1Gbps physical network links). We note this apparent discrepancy only exists at the 1% loss level, which is above our requirements. The statistics for all other packet loss levels (from 0% to 0.1%) were the same as our previous work. In the end, we discounted the significance of this difference in the 1% case and did not pursue it further.

| nuttcp "-r" times (seconds) using exponential backoff | | | | | |
|---|---|---|---|---|---|
|  | 0% | 0.001% | 0.01% | 0.1% | 1% |
| min | 2.61 | 2.61 | 2.61 | 2.68 | 2.86 |
| max | 2.61 | 2.79 | 3.44 | 3.64 | 4.34 |
| median | 2.61 | 2.61 | 2.65 | 2.81 | 3.32 |
| mean | 2.61 | 2.63 | 2.69 | 2.84 | 3.25 |
| stddev | 0.00 | 0.04 | 0.10 | 0.13 | 0.31 |
| 3 sigma | 2.61 | 2.76 | 2.98 | 3.23 | 4.18 |

Table 4. nuttcp "-r" times for crosstalk-corrected file transfers using the standard Linux retransmission timer calculations (exponential backoff), for the indicated packet loss levels. This is using 100% of the 476 Mbps of bandwidth. The 3 sigma time is the time in seconds under which 99.87% of all transfers will complete.

| nuttcp "-r" times (seconds) using new retransmission algorithm | | | | | |
|---|---|---|---|---|---|
|  | 0% | 0.001% | 0.01% | 0.1% | 1% |
| min | 2.61 | 2.61 | 2.62 | 2.71 | 2.86 |
| max | 2.61 | 2.75 | 2.82 | 3.04 | 3.33 |
| median | 2.61 | 2.61 | 2.66 | 2.82 | 3.04 |
| mean | 2.61 | 2.61 | 2.69 | 2.82 | 3.04 |
| stddev | 0.00 | 0.02 | 0.06 | 0.04 | 0.10 |
| 3 sigma | 2.61 | 2.66 | 2.87 | 2.95 | 3.35 |
| diff in 3 sigma | 0.00 | -0.10 | -0.10 | -0.28 | -0.83 |

Table 5. nuttcp "-r" times for crosstalk-corrected file transfers using the modified retransmission timer calculations as described in this paper, for the indicated packet loss levels. This is using 100% of the 476 Mbps of bandwidth. The 3 sigma time is the time in seconds under which 99.87% of all transfers will complete. The "diff in 3 sigma" shows the reduction in 3 sigma time that the modified algorithm achieves.

*E. Crosstalk-Corrected Image Transfers*

Table 6 gives some statistics for the test runs plotted in Figure 3. The nuttcp "-r" time is the time reported by the receiver-side of the nuttcp utility, and gives the duration from when the first data byte is sent to when the last byte is acknowledged. This includes an extra RTT (180ms) worth of delay because the window scaling option cannot be used for the first 65K of data [1][10]. It also includes an extra ½ RTT, waiting for the returning ACK on the last byte of data. In a production system, the application software could be designed such that the receiver would know that it has received the final byte of data without needing to wait for an explicit indication from the sender. (Note that the description of nuttcp "-r" time is from the perspective of the sender, although nuttcp is actually reporting the receiving time of those same events. This is an aid when examining the packet traces, which were taken from the sending node. The durations are identical.)

| Run ID | Packet Loss | nuttcp "-r" Time (sec) | Box Width Time (sec) | Retrans Tail Time (sec) | True Transfer Time (sec) | Total Packets Lost (avg) |
|---|---|---|---|---|---|---|
| file202 | 0% | 2.746 | 2.385 | 0 | 2.476 | 0 |
| file130 | 0.001% | 2.748 | 2.387 | 0 | 2.478 | 0.9 |
| file224 | 0.01% | 2.758 | 2.397 | 0 | 2.488 | 9 |
| file412 | 0.1% | 2.937 | 2.410 | 0.167 | 2.667 | 88 |
| file407 | 1% | 3.015 | 2.474 | 0.181 | 2.745 | 884 |

Table 6. True Transfer Times and other information for the test runs plotted in Figure 3 (including two test runs not plotted). The network has 180ms latency (RTT). These times were achieved with 95% of the 476 Mbps of bandwidth. The Run ID is a randomly-assigned identifier. The last column indicates the average number of packets expected to be lost during that transfer (each transfer requires 88,400 packets at MTU 1448). The actual number of packets lost were close to expected.

The Box Width Time is the amount of time that the data is being transmitted at the maximum rate and limited only by the network bandwidth available. The "box" nomenclature is used because the throughput plots look like a box (see Figure 3 of Freemon[1]).

The Retransmission Tail Time is the interval from the end of the Box until the last byte of data is successfully retransmitted. If there are no retransmissions sent after the Box ends, then there is no Retransmission Tail.

The True Transfer Time is the time necessary to transfer the data in a production system. This is the real figure of merit. It is derived from the nuttcp "r" Time, taking into account that the window scaling delay can be easily eliminated by just holding the TCP session open, and the receiver can know that the transfer is complete faster than what nuttcp reports.

The relationships between these values can be expressed as follows:
- $N = 1\ RTT + B + R + 1\ RTT$
  (the 1st RTT is for window scaling option to take effect, the 2nd RTT for the final ACK/FIN)
- $T = B + R + ½\ RTT$
  (the ½ RTT is the time needed for the data to travel from La Serena to NCSA)
- $T = N - (3/2)\ RTT$

where
  N is the nuttcp "-r" Time
  B is the Box Width Time
  R is the Retransmission Tail Time
  T is the True Transfer Time
  RTT is 0.180 seconds

Table 7 provides the True Transfer Times for the various levels of packet loss, and represents the time required to transfer crosstalk-corrected images in a production system. The statistics are calculated from a sample size of 160 for the higher packet loss levels. The lower packet loss levels did not require as many samples to establish the result.

It is important to note, when comparing Table 6 with Table 7, that the file transfers in Table 6 were using only 95% of the bandwidth used in Table 7, and that Table 6 is just a single instance (sample) of a crosstalk-corrected image transfer.

Table 7 is an update to Table 2 in Freemon[1] incorporating the improvements discussed in this paper, for the 180ms latency case. In comparing the two tables, it should be noted that the prior work subtracted 1 RTT from the nuttcp times, whereas we argue here, and reflect in Table 7, that the more correct adjustment to the nuttcp times is (3/2) RTT.

| | True Transfer Times (s) | | | | |
|---|---|---|---|---|---|
| | 0% | 0.001% | 0.01% | 0.1% | 1% |
| min | 2.34 | 2.34 | 2.35 | 2.44 | 2.59 |
| max | 2.34 | 2.48 | 2.55 | 2.77 | 3.06 |
| median | 2.34 | 2.34 | 2.39 | 2.55 | 2.77 |
| mean | 2.34 | 2.34 | 2.42 | 2.55 | 2.77 |
| stddev | 0.00 | 0.02 | 0.06 | 0.04 | 0.10 |
| 3 sigma | 2.34 | 2.39 | 2.60 | 2.68 | 3.08 |

Table 7. True Transfer Times on a network with 180ms latency (RTT) for the indicated packet loss levels, derived from larger sample sizes. This is using 100% of the 476 Mbps of bandwidth. The 3 sigma time is the time in seconds under which 99.87% of all transfers will complete.

V. DISCUSSION

A. Benefits

The use of HTB is significantly less complicated than designing, implementing, deploying, and supporting custom application code to actively manage bandwidth allocations at run-time. This is a static policy specification, so there are no real-time updates to bandwidth allocations (cwnd) as application events occur, such as the arrival of a new high-priority file to be transferred.

With this solution, there is no need for application-layer functionality to correctly sequence file and data transfers so they do not compete with each other. The LSST system can initiate transfers for crosstalk-corrected images, raw images, and catchup images at the same time without causing any problems. All anticipated types of traffic can be accommodated.

This solution does not need to know about or track individual sockets applications/processes or TCP sessions (which is where cwnd is set).

Ad hoc traffic originating on a replicator node receives all of the benefits described in this paper. Ad hoc traffic originating elsewhere can be routed through replicator nodes so they can benefit from these enhancements without needing to run modified Linux kernels. Indeed, they should not be running with congestion control disabled as their traffic will (presumably) be running over local shared-bandwidth network segments as well as the international network. They benefit since they can borrow any unused bandwidth on the international network, and packet loss (the main factor in poor throughput on shared-bandwidth LFNs for bulk transfers) should be significantly lower as a result of the HTB shaping.

## VI. Future Work

This work was performed in a lab by simulating network characteristics as described. Testing on a real international network is planned.

Additional work on the Linux kernel code modifications is needed before deployment into a production environment. In addition to enabling more flexible assignment of configuration parameters (eliminating some currently hard-coded values), the kernel should be changed to automatically revert back to normal TCP congestion control in response to catastrophic packet loss.

One can envision a tighter coupling of the techniques discussed in this paper with dynamic network provisioning systems such as OSCARS[7]. For example, a higher level service could control, coordinate, and synchronize between OSCARS virtual circuits and its bandwidth reservations, and HTB which is implementing the traffic shaping policies of the application system.

## VII. Conclusion

Using virtual circuits, HTB, and disabling TCP congestion control enables full utilization and throughput of expensive international networks. This brings geographically distant locations closer together, and opens up new possibilities for the use of long distance networks that were previously not practical.

Although the LSST project provided the use case scenarios discussed, this solution is a general one that can be adopted by any project or organization. This is most relevant for those entities using virtual circuits over long distances.